\documentclass[10pt,a4page]{elsarticle}
\usepackage{natbib}
\usepackage{amsfonts,mathrsfs}
\usepackage{amssymb}
\usepackage{amsmath}
\usepackage{graphicx}
\usepackage{epstopdf}
\usepackage{multirow}
\usepackage{lineno}

\begin{document}

\begin{frontmatter}

\title{\textbf{On the contribution of Rossby waves driven by surface buoyancy fluxes to low-frequency North Atlantic steric sea surface height variations}}

\author{Peter Kowalski\corref{cor1}}
\ead{p.kowalski@imperial.ac.uk}
\address{Department of Physics, Imperial College, London, UK}
\cortext[cor1]{Corresponding author}

\begin{abstract}
Previous studies have shown that wind-forced baroclinic Rossby waves can capture a large portion of low-frequency steric SSH variations in the North Atlantic. In this paper, we extend the classical wind-driven Rossby wave model derived in a 1.5 layer ocean to include surface buoyancy forcing, and then use it to assess the contribution from buoyancy-forced Rossby waves to low-frequency North Atlantic steric SSH variations. In the tropical-to-mid-latitude North Atlantic we find that wind-driven Rossby waves are dominant, however, in the eastern subpolar North Atlantic their contribution is roughly the same as that of buoyancy-forced Rossby waves, where together they capture up to 50\% of low-frequency steric SSH variations.
\end{abstract}


\end{frontmatter}

\section{Introduction}

In the North Atlantic variations in sea surface height (SSH) on interannual and longer timescales have been shown to be primarily due to variations in steric SSH (e.g. Forget and Ponte \citeyear{40}; Polkova et al., \citeyear{7}; Piecuch and Ponte \citeyear{1}; Piecuch and Ponte \citeyear{2,14}), which is defined as
\begin{equation}\label{M1}
\eta_S~=~-{1\over{\rho_0}}\int^{0}_{-D}\Delta\rho~dz,
\end{equation}
where $\eta_S$ denotes the steric SSH, $D$ is the ocean depth, and $\Delta \rho = \rho-\rho_0$ is the density anomaly with $\rho_0$ a characteristic ocean density. Many studies have thus linked low-frequency steric SSH variations in the North Atlantic to wind-forced baroclinic Rossby waves (e.g. Sturges et al. \citeyear{4}; Cabanes et al. \citeyear{39};  Zhang and Wu \citeyear{3}; Polkova et al. \citeyear{7}; Zhang et al. \citeyear{13}; Calafat et al. \citeyear{44}), which can take many years to cross an ocean basin and are therefore thought to influence climate on decadal timescales (e.g. Schneider and Miller \citeyear{21}). In particular, all previous studies combined have shown that the linear first baroclinic mode Rossby wave model forced by winds can capture a significant portion of both the phase and amplitude of steric SSH variations in the interior of the eastern North Atlantic, but that it is generally less skillful in the western part of the basin (e.g. Cabanes et al. \citeyear{39}; Zhang and Wu \citeyear{3}; Zhang et al. \citeyear{13}). With regards to buoyancy-forced Rossby waves, Piecuch and Ponte \citeyearpar{2} showed that they significantly influence the phase of the steric SSH in the tropical South Atlantic, however, their contribution to low-frequency North Atlantic steric SSH variations in the North Atlantic is yet to be explored.

The most widely used Rossby wave model of the steric SSH in the literature is the wind-driven Rossby wave model derived in a 1.5-layer ocean (e.g. Schneider and Miller, \citeyear{21}; Zhang and Wu, \citeyear{3}). In this framework the density is modelled as two layers of constant density with the interface between the two layers taken to represent the depth of the pycnocline, which is the depth in the permanent pycnocline at which the vertical density gradient is a local maximum (e.g. Feucher et al., \citeyear{12}). Furthermore, the ocean dynamics are governed by the upper ocean linear vorticity balance in a wind-driven ocean that is in geostrophic and hydrostatic balance, but with only the layer above the pycnocline depth in motion. In this paper we allow the upper layer density in the classical 1.5 layer ocean model framework to vary and subsequently extend the wind-driven Rossby wave model to include surface buoyancy forcing.

This paper is organised as follows: In Section 2 we describe the (model) data set used in this study (the ``observations'') and methodology. In Section 3 we examine the role of Rossby waves forced by surface buoyancy fluxes by comparing low-frequency SSH variations predicted by the wind-and buoyancy-forced Rossby wave model derived in a 1.5 layer ocean (e.g. Huang \citeyear{25}) with those of the ``observed''; The derivation of this model can be found in the appendix. Section 3 also includes a summary of the findings along with some suggestions for future work.

\section{Data and methodology} \label{data}

The data set that we use is the ECCO v4r1 ocean state estimate (Forget et al., \citeyear{20}) produced by the Estimating the Circulation and Climate of the Ocean (ECCO) consortium (Wunsh and Heimbach, \citeyear{11}). The data covers the period 1992-2011. Furthermore, the horizontal resolution of this particular solution in the North Atlantic is $0.25^{o}$ with 50 vertical levels of varying thickness, and the time resolution is monthly. For a detailed description of the ECCO v4r1 ocean state estimate see Liang et al. \citeyearpar{36}, who used this version of the solution to study bidecadal changes in ocean heat content. Data for all the relevant variables (e.g. surface heat fluxes, density and wind-stress) used in this study are readily available in ECCO v4r1.

The model steric SSH, denoted by $\eta_\rho$, is computed using Eq. (\ref{M1}) with $\rho_0 = 1027 \text{kg}\text{m}^{-3}$. We form time series of anomalies for $\eta_{\rho}$, and the forcing terms in the Rossby wave model, by removing the time mean, linear trend and seasonal cycle. The resulting time series of anomalies for the forcing terms are then used to derive time series of model-simulated steric SSH anomalies, which are denoted by $\eta_S$. The time series of $\eta_S$ and $\eta_{\rho}$ are then low-pass filtered to remove any signals shorter than 1 year to focus on interannual-to-decadal timescales. The time series $\eta_S$ that are the best match to $\eta_{\rho}$ are determined from the skill metric
\begin{equation} \label{Skill}
S~=~ \bigg{(}1-{<(\eta_{\rho}-\eta_S)^2>\over{<\eta_{\rho}^2>}}\bigg{)}\times 100\%.
\end{equation}
In this formula values of $S$ range from $-\infty$ to $100\%$. $S\rightarrow 100\%$ indicates that $\eta_S$ is very close to $\eta_{\rho}$ in both phase and amplitude while $S<0$ indicates that $\eta_S$ may capture the phase but overestimates the magnitude of $\eta_{\rho}$. Briefly, we found that the regions with statistically significant predictions of $\eta_{\rho}$ are roughly those with $S>10\%$, therefore, in all figures showing forecast skill [Fig. 2] the regions in which $S<10\%$ have been masked.

\section{Response of steric SSH to buoyancy-forced Rossby waves}

As shown in the appendix, in a simple 2-layer geostrophic and hydrostatic ocean with upper layer density that varies with space and time, the steric SSH can be written
\begin{equation} \label{1A}
\eta_S~=~\eta_P+\eta_L,
\end{equation}
where $\eta_L$ is the component of $\eta_S$ that is associated with local surface buoyancy fluxes and $\eta_P$ is associated with wind- and buoyancy-forced baroclinic Rossby waves. In this section we focus on the contribution of $\eta_P$ to $\eta_S$. The Rossby wave equation for $\eta_P$ is
\begin{equation}
\label{PP1}{\partial \eta_P\over{\partial t}}+C_R{\partial \eta_P\over{\partial x}}~=~-{{g'\over{g}}}W_E-{B\over{2\rho_0}}-\epsilon \eta_S,
\end{equation}
where $t, x,y$ refer to time and the distances in the east-west and north-south directions, respectively, $g'$ is the reduced gravity, $C_R$ is the Rossby wave phase speed, $W_E$ is the local Ekman pumping that is derived using the wind-stress $\boldsymbol{\tau}$ from ECCO v4r1, $B$ is a source of buoyancy, and note that we have added dissipation $\epsilon$ which represents, among other possible mechanisms, destabilization of long Rossby waves by baroclinic instability (Lacasce and Pedlosky \citeyear{38}). Following Piecuch and Ponte \citeyearpar{2} we assume that $B$ represents mixed layer fluxes of heat and freshwater, however, as in Cabanes et al. \citeyearpar{39}, we find that freshwater fluxes are negligible over the entire North Atlantic (not shown), therefore,
\begin{equation}
\label{f1}
-{B\over{\rho_0}}~=~{\alpha_T Q_{net}\over{\rho_0c_p}},
\end{equation}
where $Q_{net}$ is the surface heat flux in ECCO v4r1, $\alpha_T= 2\times 10^{-4} {\text{K}}^{-1}$ is the coefficient of thermal expansion and $c_p=4028 \text{J}{\text{Kg}^{-1}\text{K}^{-1}}$ is the specific heat capacity.

The analytical solution to Eq. (\ref{PP1}) is obtained by integrating along Rossby wave characteristics $x-C_Rt =constant$ (e.g. Cabanes et al., \citeyear{39}; Fu and Qiu, \citeyear{5}). The resulting solution after replacing $-B/\rho_0$ with Eq. (\ref{f1}) is
\begin{equation} \label{N1}
\eta_P~=~ \eta_B + F_{W_E} +F_{Q},
\end{equation}
where
\begin{eqnarray}
\label{N2}
\eta_B &=& \exp\bigg{(}\epsilon{x_E-x\over{C_R}}\bigg{)}\eta_S\bigg{(}x_E,y,t+{x_E-x\over{C_R}}\bigg{)}, \\
\nonumber
F_{W_E}&=& -{g'\over{g C_R}}\int^x_{x_E}W_E\bigg{(}x',y,t+{x'-x\over{C_R}}\bigg{)}\\
\label{N3} &\times& \exp\bigg{(}\epsilon{x'-x\over{C_R}}\bigg{)}~dx',\\
\nonumber F_{B}&=& {\alpha_T\over{2\rho_0c_pC_R}}\int^x_{x_E}Q_{net}\bigg{(}x',y,t+{x'-x\over{C_R}}\bigg{)}\\
\label{N4}&\times& \exp\bigg{(}\epsilon{x'-x\over{C_R}}\bigg{)}~dx',
\end{eqnarray}
and $\eta_S(x_E,y,t)$ is the eastern boundary steric SSH, which is computed using the ECCO model proxy for the observed steric SSH, $\eta_{\rho}$, [see Section. 2] at the eastern boundary, $x_E$. As seen in Fig. 2, the eastern is a few degrees away from the coastlines and note that it is similar to that of Zhang et al. \citeyearpar{13}. Briefly, we performed model simulations in which we varied the position of the eastern boundary and found that moving it closer to the coastlines reduces the skill of the model in the interior of the eastern North Atlantic (not shown). The reduction in skill could be due to the inclusion of SSH anomalies that are associated with other factors instead of westward propagating baroclinic Rossby waves (e.g. topography or coastally trapped waves, as suggested by Zhang et al. \citeyear{13}).

We assess the contribution of buoyancy-forced Rossby waves by comparing forecast skill [Eq. (\ref{Skill})] derived when $F_Q\neq0$ and $F_Q=0$ in Eq. (\ref{N1}). Following previous studies (e.g. Zhang and Wu \citeyear{3}; Zhang et al. \citeyear{13}), we let $g'$, $\epsilon$ and $C_R$ depend on latitude for simplicity. Estimates for these parameters are derived by finding $g'$, $\epsilon$ and $C_R$ that yield the largest values for forecast skill when $F_Q\neq 0$ and $F_Q=0$ at each grid point, and then finding the zonal average for each parameter over both experiments but only using the regions where the forecast skill is greater than 10\%. In these experiments we let $g'\in [0.01\text{m}\text{s}^{-2} \sim 0.06\text{m}\text{s}^{-2}]$ and $\epsilon\in [(0.01\text{yr})^{-1}\sim (16\text{yr})^{-1}]$ (e.g. Zhang and Wu \citeyear{3}; Zhang et al. \citeyear{13}), and for $C_R$ we first obtain an estimate at each latitude using the Lioville-Green approximation (e.g. Piecuch and Ponte \citeyear{2}),
\begin{equation} \label{LG}
C_R~=~{\beta\over{\pi f^2}}\bigg{[}\bigg{<}\int^0_{-D}N(x,y,z)\bigg{>}\bigg{]}^2,
\end{equation}
and then allow $C_R$ to vary around this estimate. In the above equation, $\beta= df/dy$, $N$ is the Brunt-Vaisala frequency computed at each grid point using the time-mean density in ECCO v4r1, and $<.>$ denotes zonal average at each latitude. Fig. 1 shows $g'$ [Fig. 1(a)], $\epsilon$ [Fig. 1(b)] and $C_R$ [Fig. 1(c)]. Note that $g'$, which is a measure of the ocean stratification, decreases roughly monotonically from the tropics to the poles; $\epsilon$ in the tropics is much larger than in the subtropical-to-subpolar North Atlantic; and $C_R$ is very similar to that obtained by Zhang \& Wu \citeyearpar{3} and Zhang et al. \citeyearpar{13}. Note further that the values for forecast skill derived by finding optimal $g'$, $\epsilon$ and $C_R$ at each grid point are similar to those that are obtained using the parameter values in Fig. 1 (not shown). Finally, similar to Chelton and Schlax \citeyearpar{17}, we find that there is a discrepancy between $C_R$ in Fig. 1(c), which is the phase speed of our observed proxy for the steric SSH, and that predicted by Eq. (\ref{LG}). In the literature this difference has been linked to factors such as inclusion of relative vorticity and Doppler shift by the depth mean flow (e.g. Tulloch et al. \citeyear{49}; Samelson, \citeyear{47}; Klocker and Marshall, \citeyear{48}), effects which are absent in Eq. (\ref{LG}) and our Rossby wave model.

\subsection{Results and discussion}

Fig. 2(a) shows the forecast skill $S_1$ when the solution for $\eta_P$ in Eq. (\ref{N1}) is determined when $F_Q=0$ and Fig. 2(b) shows forecast skill $S_2$ when $F_Q\neq0$. As in Cabanes et al. \citeyearpar{39}, the effect of eastern boundary steric SSH anomalies, which in our model can be determined by isolating $\eta_B$ in Eq. (\ref{N1}), is confined to the regions near the eastern boundary, and although we do not show figures for this result, this effect of $\eta_B$ can be easily deduced from both Fig. 2(a) and Fig. 2(b). The regions where $S_1$ and $S_2$ are greater than 10\% are generally the same with the exception of some parts of the subtropical North Atlantic. In the interior tropical-to-mid-latitude North Atlantic, it can be seen that the difference between $S_2$ and $S_1$ is relatively small, which shows that wind-forced Rossby waves dominate in this region. In the eastern subpolar North Atlantic from $55-60^{o}\text{N}$, $S_1$ and $S_2$ range from 20-30\% and 40-50\% respectively, therefore, the contribution of buoyancy-forced Rossby waves is roughly the same as that of wind-driven Rossby waves. However, note that in the western subpolar North Atlantic from $50-55^{o}\text{N}$, it can be seen by comparing $S_1$ with $S_2$ that the inclusion of surface buoyancy forcing in Eq. (\ref{N1}) has led to a small reduction in forecast skill.

Previous studies (e.g. Cabanes et al. \citeyear{39}; Zhang \& Wu, \citeyear{3}; Zhang et al. \citeyear{13}) have found that wind-driven Rossby waves cannot be used to predict steric SSH variations in the Mid-Atlantic Ridge and the western subpolar North Atlantic, specifically the Labrador Sea. We obtain the same result [Fig. 2(a)] and as seen in Fig. 2(b), buoyancy-forced Rossby waves do not enhance the predictive skill of Eq. (\ref{N1}) in these regions of the North Atlantic. In the literature low-frequency steric SSH variations in these regions of the North Atlantic have been linked to other factors and processes instead of baroclinic Rossby waves. In particular, Buckley et al. \citeyearpar{29} showed that variations in upper ocean heat content, which was shown by Forget and Ponte \citeyearpar{40} to significantly influence the steric SSH across the entire North Atlantic, can be attributed to surface heat fluxes and Ekman transport, with bolus transports and diffusion also playing a key role in the subpolar North Atlantic. In addition, Osychny and Cornillion \citeyearpar{50} showed that the Mid-Atlantic Ridge breaks the coherent structure of baroclinic Rossby waves originating on the eastern side of the subtropical North Atlantic. Finally, note that buoyancy forcing enters the Rossby wave model via a local model for the upper ocean density content [Eq. (A.10) in the appendix] and extending this model to include some of the aforementioned factors may subsequently improve the predictive skill of the Rossby wave model.

Briefly, with regards to previous studies that have investigated the role of wind-driven Rossby waves in North Atlantic steric SSH variability, our results [Fig. 2(a)] are most similar to those obtained by Zhang et al. \citeyearpar{13}, who also included damping in their wind-driven Rossby wave model, however, our values for forecast skill on the western side of the tropical-to-mid-latitude North Atlantic are generally higher. This difference may be partly due to the different data set used; Zhang et al. \citeyearpar{13} compare their model-simulated SSH with SSH from satellite altimetry, whereas we compared ours with a model steric SSH derived using data from the ECCOv4 ocean state estimate and is therefore not a pure observation.

Finally, there are many studies (e.g. Schneider and Miller, \citeyear{21}; Capotondi and Alexander, \citeyear{9}; Fu and Qiu, \citeyear{5}) that suggest wind-driven Rossby waves play a role in the low-frequency variability of the steric SSH in the North Pacific, but the role of buoyancy forced Rossby waves is yet to be explored. It would thus be interesting to conduct a similar study for the North Pacific.

\section*{Acknowledgements}

The author is grateful to Arnaud Czaja for constructive comments on an earlier version of the manuscript.

\medskip


\newpage
\appendix

\section{Linear oceanic model}

\subsection{Model framework}

We consider a two-layer rectangular basin of constant depth, $D$, and linear perturbations about a reference state of rest. The free surface is $\eta_S(x,y,t)$ and the depth of the interface is $-H_1+\eta_1$, where $-H_1$ is the depth of the interface at rest and $\eta_1(x,y,t)$ is the interface depth displacement. The upper layer density is $\rho_1(x,y,t)$ and for simplicity, we take the lower layer density to be constant, i.e. $\rho_2=\hat{\rho}_2$. Under the assumption of geostrophy and hydrostatic balance the pressure, horizontal velocity field and linear vorticity balance in the upper layer are, respectively,
\begin{eqnarray}
\label{A1} P_1 &=&-g\rho_1(z-\eta_S)\\
\label{A2} \boldsymbol{u}_1 &=&\bigg{(}{1\over{f\rho_0}}\bigg{)}(\hat{z}\times \nabla_H P_1),\\
\label{A3} {\beta\over{f}}\int^{-\delta_E}_{-H_1+\eta_1}v_1~dz &=& W_E-{\partial \eta_1\over{\partial t}},
\end{eqnarray}
and in the lower layer
\begin{eqnarray}
\label{A4} P_2&=&-\hat{\rho}_2g(z+H_1-\eta_1) - g\rho_1(-H_1+\eta_1-\eta_S), \\
\label{A5} \boldsymbol{u}_2 &=&{1\over{f\rho_0}}(\hat{z}\times \nabla_H P_2),
\end{eqnarray}
where $g$ is gravity; $\rho_0$ a characteristic ocean density; $\delta_E$ is the Ekman layer; $f$ the Coriolis parameter with $\beta$ its meridional derivative; and $W_E = \hat{\boldsymbol{z}}\cdot (\nabla_H \times [{\boldsymbol{\tau}/{\rho_0f}}])$ is the local Ekman pumping with $\boldsymbol{\tau}$ the wind-stress.

The upper layer density is decomposed as $\rho_1 = \hat{\rho}_1+\rho'_1$ where the hat variable refers to the reference state and the prime a small deviation from it ($\rho_1'\ll \hat{\rho}_1$). Combining Eq. (\ref{A1}-\ref{A3}), and then using this decomposition for $\rho_1$ and the standard approximations (e.g. $\eta_1/H_1\ll1$, $\rho_1/\rho_0\approx 1$ and thin Ekman layer) yields the linearized upper layer linear vorticity balance,
\begin{equation}
\label{v1} \bigg{(}{\beta g H_1\over{f^2}}\bigg{)}{\partial \eta_S\over{\partial x}}+\bigg{(}{H_1^2\beta g\over{2f^2\rho_0}}\bigg{)}{\partial \rho_1'\over{\partial x}}~=~ W_E-{\partial \eta_1\over{\partial t}}.
\end{equation}
Similarly, Eq. (\ref{A4}) can be combined with Eq. (\ref{A5}) giving the following expression for the lower layer horizontal velocity field:
\begin{equation}
\label{u2}
 \boldsymbol{u}_2 ~=~ \bigg{(}{g'\over{f}} \bigg{)}(\hat{\boldsymbol{z}}\times \nabla_H \eta_1)+\bigg{(}{g\over{f}}\bigg{)}(\hat{\boldsymbol{z}}\times \nabla_H \eta_S)+\bigg{(}{gH_1\over{f\rho_0}}\bigg{)}(\hat{\boldsymbol{z}}\times \nabla_H \rho_1'),
\end{equation}
where $g'~=~{g(\hat{\rho}_2-\hat{\rho}_1)/{\rho_0}}$ is the reduced gravity. As in the classical 1.5 layer ocean we assume that the lower layer is at rest ($u_2=0$), which leads to the following relation for $\eta_S$:
\begin{equation}
\label{S1}
\eta_S~=~-{g'\over{g}}\eta_1 -{H_1\over{\rho_0}}\rho_1'.
\end{equation}
Note that it can easily be shown using Eq. (\ref{M1}) that in our model framework $\eta_S$ is the steric SSH. The time derivative of $\eta_S$ is
\begin{equation}
\label{S1a}
{\partial \eta_S \over{\partial t}}~=~-{g'\over{g}}{\partial \eta_1\over{\partial t}} -{H_1\over{\rho_0}}{\partial\rho_1'\over{\partial t}}.
\end{equation}
Now, since the reference state is at rest, there are no background horizontal density gradients that can generate upper layer density anomalies ($\rho_1'$) and the only physical mechanism left to do so is a surface buoyancy flux, $B$. Therefore,
\begin{equation} \label{A4*}
{\partial (H_1\rho_1')\over{\partial t}}~=~B,
\end{equation}
where $H_1\rho_1'$ is the upper ocean density content, and then substituting this equation into Eq. (\ref{S1a}) gives
\begin{equation} \label{A6}
{\partial \eta_S\over{\partial t}}~=~-{g'\over{g}}{\partial \eta_1\over{\partial t}}-{B\over{\rho_0}}.
\end{equation}
We show next that $\eta_S$ can be written as the sum of a local component due to local surface buoyancy forcing, and a dynamical component that is associated with wind- and buoyancy-forced Rossby waves.

\subsection{Local and Rossby wave steric SSH dynamics} \label{Deriv}

The key step is to write the upper layer linear vorticity balance [Eq. (\ref{v1})] as
\begin{equation}\label{v1*}
 \bigg{(}{\beta g H_1\over{f^2}}\bigg{)}{\partial \eta_P\over{\partial x}}~=~ W_E-{\partial \eta_1\over{\partial t}},
\end{equation}
where
\begin{equation}\label{P2*}
\eta_P~=~\eta_S+{H_1\over{2\rho_0}}\rho_1'
\end{equation}
is a sea surface height, and the subscript P here is for pressure since $\eta_P$ arises from the vertically integrated meridional velocity [l.h.s of Eq. (\ref{A3})]. Differentiating this equation and using Eq. (\ref{A4*}) gives
\begin{equation}\label{A8}
{\partial \eta_P\over{\partial t}}~=~{\partial \eta_S\over{\partial t}}+{B\over{2\rho_0}}.
\end{equation}
Substituting Eq. (\ref{A6}) into the above equation gives
\begin{equation}\label{A9}
{\partial \eta_P\over{\partial t}}~=~-{g'\over{g}}{\partial \eta_1\over{\partial t}}-{B\over{2\rho_0}},
\end{equation}
and then combing this equation with Eq. (\ref{v1*}) leads to the following wind- and buoyancy-forced Rossby wave equation for $\eta_P$:
\begin{equation}\label{A10}
{\partial \eta_P\over{\partial t}}+C_R{\partial \eta_P\over{\partial x}}~=~-{1\over{2\rho_0}}B-{g'\over{g}}W_E,
\end{equation}
where
\begin{equation}\label{A11}
C_R~=~-{\beta g' H_1\over{f^2}}<0,
\end{equation}
is the Rossby wave phase speed. The equivalent version of this equation in a continuously stratified ocean was  derived by Piecuch \& Ponte \citeyearpar{2} using normal modes decomposition.

From Eq. (\ref{A8}) it follows that
\begin{equation} \label{A12}
\eta_S~=~\eta_P+\eta_L,
\end{equation}
where $\eta_L$ is determined from
\begin{equation} \label{A13}
{\partial \eta_L\over{\partial t}}~=~ -{B\over{2\rho_0}},
\end{equation}
and is therefore the component of $\eta_S$ that is associated with local surface buoyancy forcing. Note also that from Eq. (\ref{A9}) the interface depth displacement $\eta_1$ can also be similarly written as
\begin{equation} \label{A15}
\eta_1~=~ {g\over{g'}}(-\eta_P+\eta_L).
\end{equation}
Finally, it is straightforward to see that when $B=0$, Eq. (\ref{A8}-\ref{A10}) reduce to the following set of equations:
\begin{eqnarray}
\label{A15}
\eta_S&=& \eta_P, \\
\label{A16}
\eta_1 &=&  -{g\over{g'}}\eta_P, \\
\label{A17}
{\partial \eta_P\over{\partial t}}+C_R{\partial \eta_P\over{\partial x}}&=&-{g'\over{g}}W_E.
\end{eqnarray}
These equations are equivalent to those derived in the classical 1.5-layer ocean configuration.

\bibliographystyle{Else}
\bibliography{SSHNew}

\renewcommand{\thefigure}{\arabic{figure}}
\renewcommand{\figurename}{Fig.}

\clearpage

\newpage

\begin{figure*}[t]
\begin{center}
\hspace*{-6mm}
\includegraphics[width=0.7\textwidth,height=0.7\textwidth]{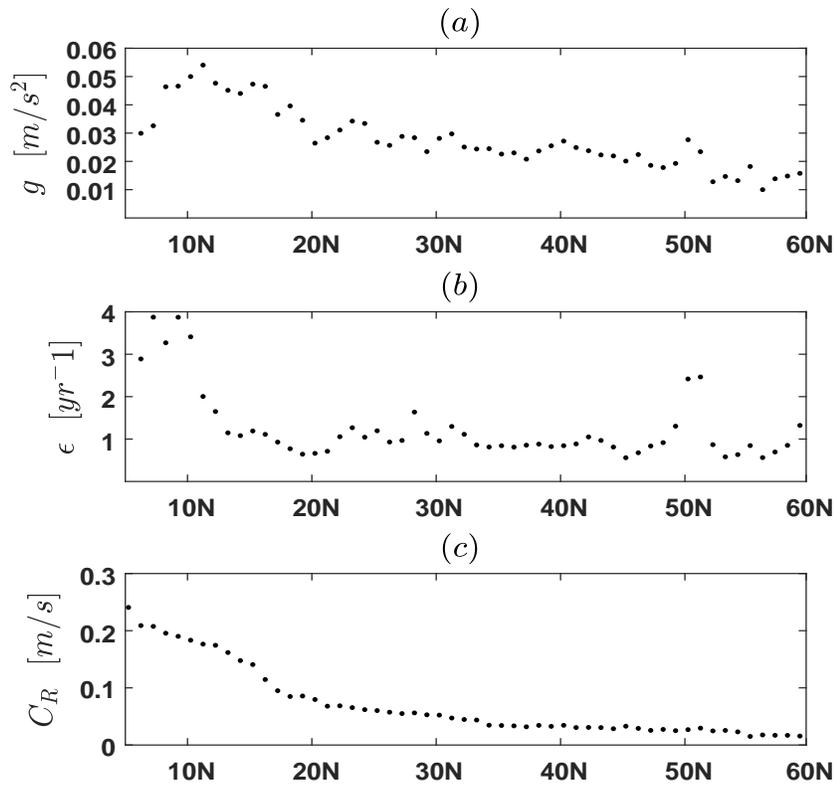}\\
\end{center}
\caption{(a) shows reduced gravity, $g'$, (b) is the dissipation rate, $\epsilon$, and (c) is the Rossby wave phase speed $C_R$}
\label{fig1}
\end{figure*}

\begin{figure*}[t]
\begin{center}
\hspace*{-6mm}
\includegraphics[width=0.8\textwidth,height=0.8\textwidth]{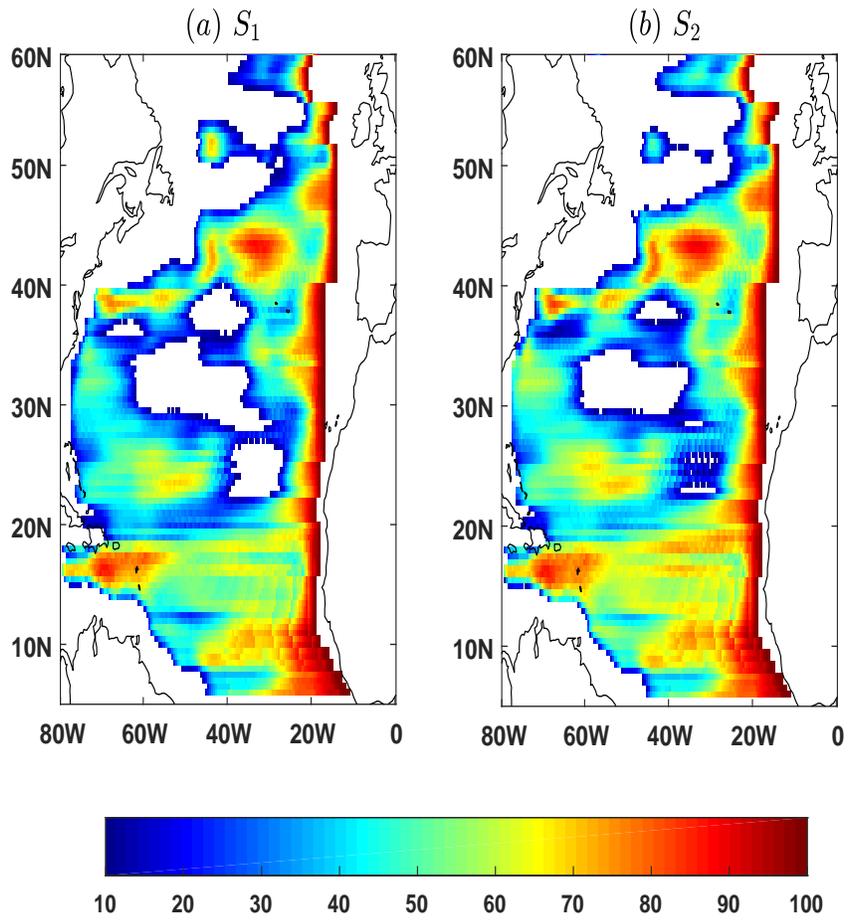}\\
\end{center}
\caption{(a) and (b) show forecast skill when $F_Q=0$ and $F_Q\neq 0$ in Eq. (\ref{N1})}
\label{fig2}
\end{figure*}


\end{document}